\documentclass[conference]{IEEEtran}
\IEEEoverridecommandlockouts
\usepackage{cite}
\usepackage{amsmath,amssymb,amsfonts}
\usepackage{algorithmic}
\usepackage{graphicx}
\usepackage{textcomp}
\usepackage{placeins}
\usepackage{xcolor}
\usepackage{flafter}
\usepackage{balance}
\usepackage{multirow}
\usepackage{subcaption} 
\usepackage[hyphens]{url}
\usepackage{booktabs}
\def\BibTeX{{\rm B\kern-.05em{\sc i\kern-.025em b}\kern-.08em
    T\kern-.1667em\lower.7ex\hbox{E}\kern-.125emX}}
\begin{document}

\title{Design Space Exploration of Backside Clock Meshes for 2\,nm GAAFET BSPDN Technology\\
}

\author{\IEEEauthorblockN{Wajid Ali}
\IEEEauthorblockA{\textit{UC Santa Cruz} \\
Santa Cruz, CA, USA \\
wali2@ucsc.edu}
\and
\IEEEauthorblockN{Muhammad Hadir Khan}
\IEEEauthorblockA{\textit{UC Santa Cruz} \\
Santa Cruz, CA, USA \\
mkhan33@ucsc.edu}
\and
\IEEEauthorblockN{Dalton Gaddy}
\IEEEauthorblockA{\textit{UC Santa Cruz} \\
Santa Cruz, CA, USA \\
dgaddy@ucsc.edu}
\and
\IEEEauthorblockN{Matthew Guthaus}
\IEEEauthorblockA{\textit{UC Santa Cruz} \\
Santa Cruz, CA, USA \\
mrg@ucsc.edu}
}

\maketitle

\begin{abstract}
Clock meshes are used in high-performance VLSI designs to minimize skew and tolerate on-chip variation, but they spend scarce routing resources on premium metal layers. Backside power delivery creates a new option: it adds thick, low-resistance metal layers on the back of the wafer, and the power grid does not consume all of them. Flip-flops remain on the frontside; a backside mesh therefore cannot drive them directly, and every connection passes through a through-silicon via. We present the first design-space exploration of backside clock meshes, implemented in OpenROAD on GT2N, a 2\,nm nanosheet technology. Four benchmarks (1,938 to 15,311 flip-flops) are explored with multi-objective Bayesian optimization, and every design point is verified by transistor-level SPICE simulation, since the cyclic mesh cannot be evaluated by static timing analysis. Across all four designs, the backside mesh consistently outperforms an identical frontside mesh, with on average 45\% lower skew, 25\% lower sink slew, 4.5\% lower power, and 28\% less frontside clock wiring, and its skew spread under 10{,}000-sample Monte Carlo is a third of the frontside mesh's.
\end{abstract}

\section{Introduction}
\label{sec:intro}

\begin{figure}[!t]
\centering
\includegraphics[width=\columnwidth]{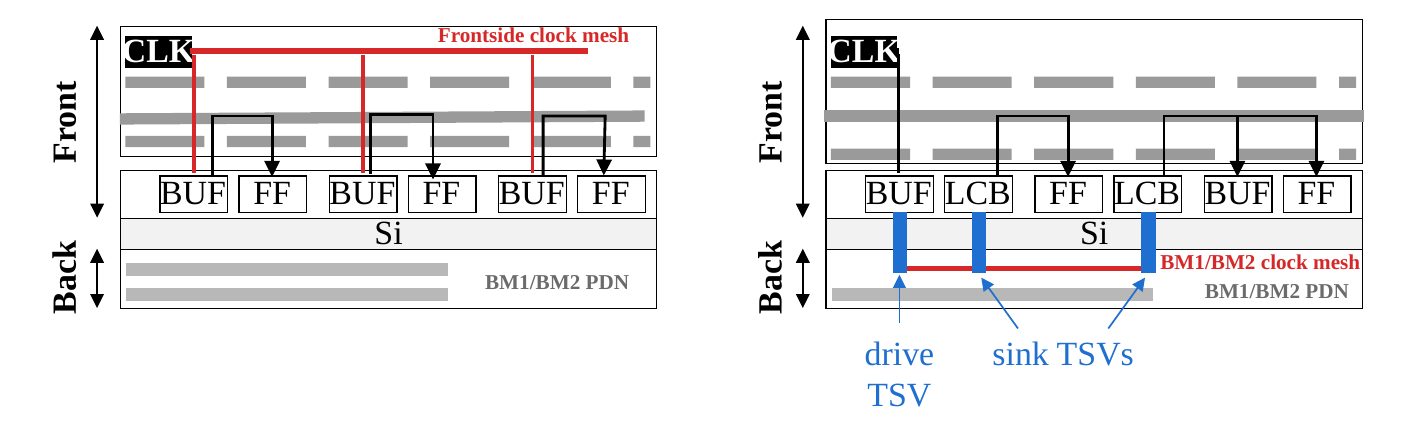}
\caption{Moving the mesh to the backside forces every mesh-to-sink connection through a TSV and cuts a buried power rail at every landing.}
\label{fig:crosssection}
\end{figure}
A clock mesh consists of a grid of wires distributed across the die and driven from multiple points. Since thousands of sink paths share the same grid, local delay differences are averaged out instead of accumulating along individual paths. This results in very low skew and good tolerance to process and environmental variations, which is why clock meshes are used in high-performance designs~\cite{restle,rajaram,guthaus2010,guthaus2013}. There also has been research efforts regarding their fast analysis for optimization~\cite{gatmesh, khan:mesh-glsvlsi}. One of the main drawbacks in clock meshes are power and routing overhead: the entire grid capacitance must be charged every cycle, while the grid also occupies routing tracks on the metal layers that are needed for signal routing and the power grid.

The power grid is itself a heavy consumer of those metal layers, and backside power delivery (BSPDN) removes it from the frontside entirely. Buried power rails beneath the standard cells are fed from a power grid on the back of the
wafer, eliminating the frontside power straps and improving IR drop at the same time~\cite{prasadbspdn,chen2021,veloso2023}. The approach is now in production at Intel and is on the TSMC and Samsung roadmaps~\cite{shamanna2023,tsmca16,samsungsf2z}. More importantly for a clock mesh, BSPDN adds backside metal layers that are
wider and thicker than the frontside layers a mesh would normally occupy, and the power grid does not use all of their tracks. In the 2\,nm GT2N technology we target, the backside metals have $3.5\times$ lower resistance per micron than the M6--M9 class and $22\times$ lower than M4/M5 (Section~\ref{sec:technology}). This low wire resistance is what makes a backside clock mesh worth investigating.

Prior work has begun to exploit backside metal for clocking, but majorly for clock \emph{trees}. Veloso et al.\ route the global H-tree portion of a high-performance CPU's clock on the backside metals and report a clear latency reduction from even that partial move~\cite{veloso2023}; Bethur et al.\ select which tree nets to flip by fanout or by a GNN-predicted timing criticality~\cite{bethur,bethurdac24}; Vanna-iampikul et al.\ share the backside metals between the power grid and selected clock trunk nets~\cite{vannaiampikul24}; Jiang et al.\ synthesize the double-sided tree directly with concurrent buffer and nTSV insertion~\cite{jiang25}; and CSCO allocates the limited backside tracks jointly between clock and signal nets~\cite{csco26}. To our knowledge, backside clock meshes have not yet been explored.

A mesh cannot simply be relocated (Fig.~\ref{fig:crosssection}). The backside layers already carry the power grid, so the mesh must fit around it. The flip-flops remain on the frontside, so every mesh-to-sink connection must cross the wafer through a through-silicon via (TSV). Each TSV passes through a buried power rail, which must be cut around it; and the shorted grid is a cyclic network that static timing analysis cannot evaluate, so every candidate configuration must be simulated at the transistor level. The question is whether the mesh still wins once the cost of reaching it from the frontside is paid.

This paper makes the following contributions:
\begin{itemize}
  \item Upstream OpenROAD support for backside routing, buried-power-rail
        cutting at TSV landings, and backside IR analysis, which the flow of
        this paper builds on.
  \item The first construction flow for backside clock meshes, implemented in
        OpenROAD on the open-source GT2N 2\,nm nanosheet BSPDN technology.
  \item A multi-objective Bayesian design-space exploration over four
        mesh parameters, with every design point verified by full-network
        HSPICE simulation.
  \item A comparison against two frontside mesh baselines on four benchmarks
        (1,938 to 15,311 flip-flops), with 10,000-sample Monte
        Carlo analysis.
\end{itemize}

Across all four designs, the evaluated backside mesh configurations achieve lower clock skew, lower maximum sink slew, and lower frontside clock wirelength than the corresponding frontside mesh configurations, and are markedly less sensitive to variation. Section~\ref{sec:flow} describes the technology and the construction flow, Section~\ref{sec:dse_method} the exploration method, Section~\ref{sec:setup} the setup, and Section~\ref{sec:results} the results.

\FloatBarrier
\section{Backside Clock Mesh Synthesis}
\label{sec:flow}
The challenges outlined above affect every step of constructing a backside clock mesh. This section first describes the target technology and then presents the construction flow: placing the mesh on the backside tracks with constrained resources after power grid construction, crossing between the backside and frontside through TSVs and buried power rails, and connecting the mesh to the frontside sinks through local clock buffers.

\subsection{Target Technology}
\label{sec:technology}
We build on GT2N, an open-source 2\,nm nanosheet technology with buried power rails and backside power delivery~\cite{gt2npdk}. GT2N provides fourteen frontside metal layers (M0--M13), tungsten buried power rails beneath the standard cells, and four backside metals: BM1/BM2 at 56\,nm width and ultra-thick BM3/BM4 at 360\,nm~\cite{gt2npdk}. The low resistance of the backside metals makes them well suited to clock meshes: based on the PDK's extracted wire resistances, BM1/BM2 are $3.5\times$ less resistive per micron than the M6--M9 class (7.5 vs.\ 26\,$\Omega$/\textmu m) and $22\times$ less resistive than M4/M5. Crossings between the frontside and backside are implemented using TSV cells placed in the standard-cell rows and modeled as a 6\,$\Omega$ series resistance~\cite{bethur}. The cell library provides two nanosheet widths (13 and 31\,nm) and five threshold-voltage options.

\subsection{Mesh Construction over a Backside PDN}
\label{sec:mesh_construction}
Power reaches the standard cells through buried power rails beneath each cell row, which are fed by a power grid on BM1 and BM2. We use these same layers for the clock mesh, placing horizontal wires on BM2 and vertical wires on BM1 in the tracks left available by the power grid. The power grid is built first, and mesh wires that overlap or are too close to power straps are shifted to the nearest clear track or removed if no valid track exists. Each remaining intersection receives a via, forming a single electrically shorted mesh net.

\subsection{Clocking Architecture and Buffering}
\label{sec:clocking_arch}
\begin{figure}[t]
\centering
\includegraphics[width=\columnwidth]{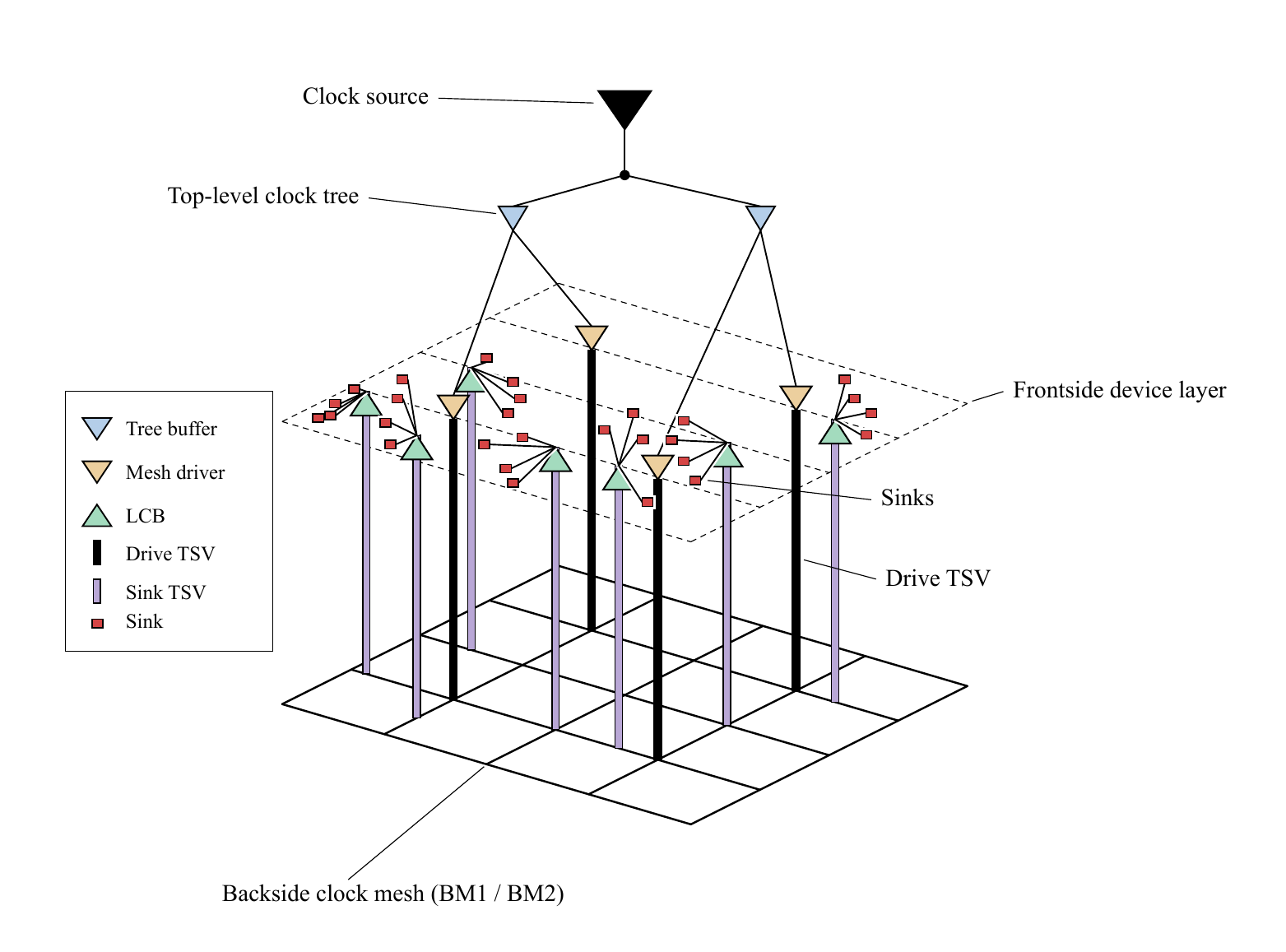}
\caption{The clock passes through three buffering stages (tree, mesh drivers, LCBs) and crosses between frontside and backside through drive TSVs and sink TSVs.}
\label{fig:clocking_arch}
\end{figure}
The mesh itself is passive; the clock reaches it and continues to the sinks through four active stages (Fig.~\ref{fig:clocking_arch}). A conventional buffer tree synthesized by TritonCTS~\cite{tritoncts}  distributes the clock from the root to the mesh driver inputs. Each driven grid intersection holds one mesh driver, a standard buffer placed on the frontside directly beneath its grid node and snapped to the nearest cell row, whose output crosses to the backside through a TSV (Section~\ref{sec:tsv}). Drivers occupy every other intersection in a
checkerboard, which halves the driver and TSV count without changing the wire grid. Below the mesh, sink connections return to the frontside through per-cluster TSVs and local clock buffers (LCBs, Section~\ref{sec:lcb}). The tree needs no precise balancing: it drives only tens to a few hundred driver inputs, and the shorted mesh averages whatever arrival differences it delivers.

All drivers on a mesh use the same buffer cell, whose drive strength is an explored parameter (Table~\ref{tab:knobs}). The exploration consistently prefers small drivers (x1--x4), even on the largest design. The LCB tier (Section~\ref{sec:lcb}) is sized independently; its role is to regenerate the clock edge for a small group of flip-flops.



\subsection{TSV Insertion and Buried Power Rail Breaking}
\label{sec:tsv}
Two kinds of TSV are used: a drive TSV at each mesh driver and a sink TSV at each LCB (on Ibex at a 5\,\textmu m pitch, 41 and 382 respectively). A clock TSV uses the same via-stack structure that connects the buried power rails to the backside grid, but carries the clock net through the BPR layer. The rail there would short to it and violate the BPR spacing rule, so the flow cuts the
rail at every landing, removing a piece sized by the TSV footprint plus a 0.224\,\textmu m clearance on each side, and deletes any rail fragments left floating between nearby cuts; on the largest design this amounts to about 2{,}500 cuts. Power delivery is unaffected, since each cut is well under a micron and the remaining rail pieces are still fed through their own via stacks. Cells displaced by the cuts are re-legalized. None of this is needed for the frontside meshes compared in Section~\ref{sec:results}.

\subsection{Local Clock Buffers and Sink Clustering}
\label{sec:lcb}
Because the flip-flops are frontside cells, giving each its own TSV would require thousands of crossings. Instead, flip-flops are grouped into clusters of at most $F_{\text{fanout}}$ sinks: candidate taps are generated along the mesh wires, and each flip-flop joins the nearest tap whose cluster has not reached the fanout bound $F_{\text{fanout}}$. Each cluster receives one LCB, legalized with its
neighbors, and one sink TSV to the LCB input; the LCB output drives its flip-flops as an ordinary routed net. On Ibex with $F_{\text{fanout}}=6$ this yields 382 LCBs for 1{,}938 flip-flops. The tier also limits the mesh load to a
few hundred buffer inputs. For the frontside baselines in Section~\ref{sec:results}, the same tier connects each LCB to the nearest point
on the frontside mesh without TSVs, and a second variant omits the tier and attaches flip-flops directly to the mesh wires.

\FloatBarrier
\section{Design-Space Exploration}
\label{sec:dse_method}

\begin{table}[!htb]
\caption{Four parameters interact enough that a per-design joint search is required.}
\label{tab:knobs}
\centering
\begin{tabular}{ll}
\toprule
Parameter & Values \\
\midrule
Mesh pitch (\textmu m)          & 1.6--14 (per-design range) \\
Mesh driver size                & x1, x2, x3, x4, x6, x8, x10, x12 \\
LCB fanout bound $F_{\text{fanout}}$ & 6, 8, 16, 24, 40, 80 \\
LCB size                        & x1, x2, x3, x4, x6, x8, x10, x12 \\
\bottomrule
\end{tabular}
\end{table}

The flow leaves four parameters open (Table~\ref{tab:knobs}), which define a few thousand configurations per design, and each evaluation requires the full flow of Section~\ref{sec:flow} plus HSPICE simulation, taking 10--50 minutes per point. Exhaustive search is out of reach, and one-at-a-time sweeps are misleading because the parameters interact: whether a 10\,\textmu m pitch is usable depends on the LCB fanout bound, since small clusters keep the sparse mesh's edges sharp while large ones push them past the slew limit.

We therefore search all four parameters jointly with the multi-objective Tree-structured Parzen Estimator (MOTPE)~\cite{tpe,motpe} in
Optuna~\cite{optuna}, minimizing worst-case sink skew and clock network power, following the use of Bayesian tuning in physical design by AutoTuner, FIST, AutoDMP, and BOOM-Explorer~\cite{metrics21,fist,autodmp,boomexplorer}. TPE handles our categorical parameters (buffer cells, discrete pitches) natively, and its multivariate mode learns which combinations work rather than scoring
each parameter independently~\cite{autodmp}. 

Because skew and power trade against each other, we retain the full non-dominated set. A trial is reported as infeasible when detailed routing fails, when any flip-flop is missing or fails to toggle, or when the worst sink slew exceeds 10\% of the clock period; infeasible outcomes still steer the sampler away from those regions.

\section{Experimental Setup}
\label{sec:setup}

\begin{figure*}[!t]
\centering
\includegraphics[width=\textwidth]{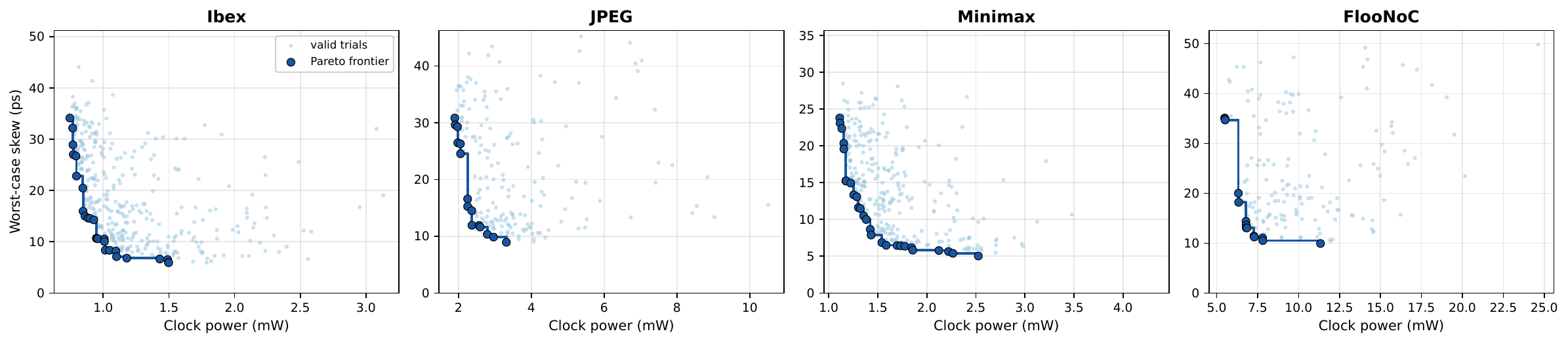}
\caption{Skew--power Pareto frontiers of the backside clock mesh. Light points are all feasible trials; dark points form the non-dominated set.}
\label{fig:pareto}
\end{figure*}

\begin{table*}[!t]
\caption{The backside mesh wins on skew, slew, power, and frontside wire on every design at a small latency cost. BS+LCB = backside mesh + LCB tier; FS+LCB = frontside mesh + LCB tier; FS Direct = frontside mesh with direct flip-flop attachment. Checkerboard mesh buffers; full-network SPICE.}
\label{tab:comparison}
\centering
\footnotesize
\setlength{\tabcolsep}{3.6pt}
\begin{tabular}{|c|l|ccc|ccc|ccc|ccc|}
\hline
\multicolumn{2}{|c|}{\multirow{2}{*}{\textbf{Benchmark}}}
 & \multicolumn{3}{c|}{\textbf{Ibex}} & \multicolumn{3}{c|}{\textbf{JPEG}}
 & \multicolumn{3}{c|}{\textbf{Minimax}} & \multicolumn{3}{c|}{\textbf{FlooNoC}} \\
\cline{3-14}
\multicolumn{2}{|c|}{}
 & \textbf{BS+LCB} & FS+LCB & FS Direct & \textbf{BS+LCB} & FS+LCB & FS Direct
 & \textbf{BS+LCB} & FS+LCB & FS Direct & \textbf{BS+LCB} & FS+LCB & FS Direct \\
\hline
\multirow{5}{*}{\rotatebox[origin=c]{90}{\scriptsize\textbf{Mesh Config.}}}
 & Mesh grid           & $9{\times}9$ & $9{\times}9$ & $9{\times}9$
                       & $25{\times}25$ & $25{\times}25$ & $25{\times}25$
                       & $9{\times}9$ & $9{\times}9$ & $9{\times}9$
                       & $20{\times}20$ & $20{\times}20$ & $20{\times}20$ \\
 & Pitch (\textmu m)   & 5 & 5 & 5 & 3.2 & 3.2 & 3.2 & 4 & 4 & 4 & 6 & 6 & 6 \\
 & Mesh buffers        & 41 & 41 & 41 & 313 & 313 & 313 & 41 & 41 & 41 & 200 & 200 & 200 \\
 & Local clock buffers & 382 & 382 & N/A & 790 & 790 & N/A & 135 & 135 & N/A & 2,611 & 2,611 & N/A \\
 & \#TSVs              & 423 & N/A & N/A & 1,103 & N/A & N/A & 176 & N/A & N/A & 2,811 & N/A & N/A \\
\hline
\multirow{5}{*}{\rotatebox[origin=c]{90}{\scriptsize\textbf{Metric}}}
 & Clock skew (ps)     & \textbf{6.80} & 15.28 & 29.78 & \textbf{9.85} & 12.34 & 19.61
                       & \textbf{20.36} & 40.00 & 39.4 & \textbf{11.05} & 25.76 & 52.27 \\
 & Max sink slew (ps)  & \textbf{14.1} & 19.1 & 53.2 & \textbf{19.9} & 23.8 & 57.8
                       & \textbf{47.5} & 76.4 & 125.6$^{\dag}$ & \textbf{23.6} & 29.7 & 111.3 \\
 & Clock power (mW)    & 1.179 & 1.261 & 0.806 & 2.961 & 3.006 & 2.088
                       & 1.153 & 1.220 & 1.053 & 7.804 & 8.184 & 5.256 \\
 & FS clock WL (\textmu m) & \textbf{2,159} & 3,055 & 3,031 & \textbf{6,008} & 10,025 & 9,017
                       & \textbf{1,896} & 2,476 & 2,753 & \textbf{17,678} & 22,214 & 23,995 \\
 & Clock latency (ps)  & 51.5 & 49.3 & 57.0 & 69.5 & 60.1 & 70.4
                       & 56.8 & 58.5 & 88.9 & 92.0 & 88.7 & 112.3 \\
\hline
\multicolumn{14}{l}{\footnotesize $^{\dag}$violates the slew budget (10\% of the clock period).}
\end{tabular}
\end{table*}

\begin{table}[!htb]
\caption{Benchmark designs: two RISC-V cores, a JPEG encoder, and a network-on-chip, from 1,938 to 15,311 flip-flops.}
\label{tab:designs}
\centering
\begin{tabular}{llrlc}
\toprule
Design & Insts & FFs & Core (\textmu m) & $T_{\mathrm{clk}}$ (ns) \\
\midrule
Minimax & 25,071  & 2,251  & $35.6 \times 35.6$   & 0.78 \\
Ibex    & 24,117  & 1,938  & $44.0 \times 43.9$   & 1.00 \\
JPEG    & 104,562 & 4,384  & $80.6 \times 80.5$   & 1.00 \\
FlooNoC & 122,079 & 15,311 & $117.0 \times 116.9$ & 1.15 \\
\bottomrule
\end{tabular}
\end{table}

All experiments run in OpenROAD with our clock-mesh extensions on the GT2N
technology of Section~\ref{sec:technology}. GT2N provides libraries and device
models at a single corner (TT, 0.7\,V, 25\,$^\circ$C), so nominal results are
reported at that corner; every skew, slew, and power number comes from
transistor-level HSPICE simulation of the complete clock network.

We evaluate four benchmarks (Table~\ref{tab:designs}): the Ibex RISC-V core
and a JPEG encoder from OpenROAD Flow Scripts~\cite{orfs}, and the Minimax
bit-serial RISC-V CPU and the FlooNoC network-on-chip from the HighTide
suite~\cite{hightide}. Each is synthesized and placed with the standard
OpenROAD flow before clock construction, so sink locations are identical for
every clock network compared on a design. Backside power straps are as wide as
the design rules allow and numerous enough to carry each design's current,
verified with OpenROAD's IR analysis. The exploration of
Section~\ref{sec:dse_method} runs 300--500 trials per design, three in
parallel, with completed configurations cached; most infeasible trials fail
the slew bound at sparse pitches.

\subsection{Variation Model}
\label{sec:variation_model}

\begin{table}[!htb]
  \caption{Monte Carlo variation parameters (Gaussian, $1\sigma$ listed). Device $\sigma$ follow N3 nanosheet projections~\cite{karner} and wire $\sigma$ follow~\cite{prasad17}; nominal values are GT2N's~\cite{gt2npdk}.}
  \label{tab:mc-params}
  \centering
  \begin{tabular}{llcl}
  \toprule
   Parameter & $\mu$ & $\sigma$ & From \\
    \midrule
    $V_{th}$        & 177\,mV          & 20 / 15\,mV$^{*}$   & \cite{karner} \\
    $L_{g}$ (gate length)        & 14\,nm           & 0.167\,nm           & \cite{karner} \\
    $W_{NS}$ (sheet width)         & 31\,nm           & 0.167\,nm           & \cite{karner} \\
    $T_{NS}$  (sheet thickness)      & 5\,nm            & 0.133\,nm           & \cite{karner} \\
    $V_{dd}$        & 0.7\,V           & 18\,mV$^{\S}$       & IR drop \\
    Wire $R$ (FS)   & extracted        & 12.5\%$^{\P}$       & \cite{prasad17,gall2020} \\
    Wire $R$ (BS)   & extracted        & 6.25\%$^{\P}$       & scaled \\
    Wire $C$        & extracted        & 3\%                 & \cite{prasad17} \\
    \bottomrule
    \end{tabular}

    \vspace{2pt}
    {\scriptsize
    $^{*}$pMOS\,/\,nMOS. $^{\S}$Worst-case backside IR drop taken as a $3\sigma$
    bound. $^{\P}$Frontside $\sigma$ is above the 5.6--9.9\% measured at
    11\,nm~\cite{prasad17} because narrower 2\,nm wires are more sensitive to
    surface and grain-boundary scattering~\cite{gall2020}; no backside data is
    published, so the wider 56\,nm backside lines are assigned half the
    frontside value.\par}
\end{table}

Monte Carlo runs 10{,}000 HSPICE samples per network on Ibex under the
parameters of Table~\ref{tab:mc-params}. Device $\sigma$ follow N3 nanosheet
projections~\cite{karner} and wire $\sigma$ follow~\cite{prasad17}; nominal
values are GT2N's, and every device and wire element is drawn independently.

\section{Results}
\label{sec:results}
\subsection{Pareto Frontiers}
\label{sec:pareto}
Fig.~\ref{fig:pareto} shows the skew--power frontier found on each design. All four share the same shape: skew falls steeply as power rises from the sparsest feasible pitch. The representative points in Table~\ref{tab:comparison} are taken at the knee, and each frontside baseline is built with the identical grid, pitch, and driver configuration.

\subsection{Backside vs.\ Frontside}
\label{sec:comparison}
Against the frontside mesh with the same LCB tier (FS+LCB), the backside mesh lowers worst-case skew on every design, by 20\% on JPEG and by 49--57\% on the other three, 45\% on average, and sharpens the worst sink slew by 16--38\%. It does so at 1.5--6.5\% lower clock power, thanks to the backside grid's lower resistance, and with 20--40\% less frontside clock wirelength, because the grid itself has left the frontside. Latency rises by 2--10\,ps on three designs, the cost of the extra TSV and LCB stage, and falls on Minimax. The direct-attach frontside mesh (FS Direct) draws the least power but pays for it in slew: its worst sink edge is
2.6--4.7$\times$ slower than the backside mesh and exceeds the 10\% slew budget on Minimax, so it is not a usable design point at these pitches.

\subsection{Power Delivery}
\label{sec:ir}
Sharing BM1/BM2 with the clock mesh and cutting the buried power rails at the TSV sites keeps the supply within budget. On Ibex, worst-case VDD drop is 30.7\,mV for the backside configuration against 32.2\,mV for the frontside mesh, and VSS drop rises from 16.8 to 22.5\,mV; both are within the budget used to size the straps in Section~\ref{sec:setup}.

\subsection{Skew Under Variation}
\label{sec:variation}

\begin{table}[!htb]
   \caption{Monte Carlo clock skew on Ibex (10{,}000 HSPICE samples per network, parameters from Table~\ref{tab:mc-params}).}
    \label{tab:mc-skew}
    \centering
    \begin{tabular}{lcccc}
    \toprule
    & BS+LCB & FS+LCB & FS Direct & CTS tree \\
    \midrule
    Nominal skew (ps)                 & \textbf{6.77}  & 15.20 & 29.82 & 32.43 \\
    $\mu_{\text{skew}}$ (ps)          & \textbf{7.13}  & 15.44 & 30.02 & 34.40 \\
    $\sigma_{\text{skew}}$ (ps)       & \textbf{0.34}  & 0.99  & 0.78  & 1.79  \\
    $\mu + 3\sigma$ (ps)              & \textbf{8.14}  & 18.43 & 32.35 & 39.77 \\
    Max skew (ps)                     & \textbf{8.57}  & 19.24 & 32.82 & 42.44 \\
    \bottomrule
    \end{tabular}
\end{table}

Table~\ref{tab:mc-skew} reports Monte Carlo results on Ibex, with a conventional TritonCTS~\cite{tritoncts} tree added as a fourth network. The backside mesh has the smallest mean skew and the smallest spread: its $\sigma_{\text{skew}}$ of 0.34\,ps is a third of the frontside mesh's and a fifth of the tree's.

\section{Conclusion}
\label{sec:conclusion}
We have presented a construction flow and design-space exploration for clock meshes on the backside metals of a BSPDN technology. Building the mesh around the backside power grid, reaching it through TSVs that cut the buried power rails, and returning to the frontside sinks through a local clock buffer tier are all workable in OpenROAD on GT2N. With every design point verified in SPICE, the backside mesh beats an identical frontside mesh on all four benchmarks in skew, slew, power, and frontside wiring, and holds that lead under process and supply variation.
\clearpage
\balance

\end{document}